\documentclass[%
 reprint,
 amsmath,amssymb,
 aps,prl,floatfix,
]{revtex4-1}

\usepackage{graphicx}
\usepackage{dcolumn}
\usepackage{bm}

\usepackage[english]{babel}
\usepackage[skip=0pt, indent=15pt]{parskip}
\usepackage{amsmath}
\usepackage{graphicx}
\usepackage[colorlinks=true, allcolors=blue]{hyperref}
\usepackage{setspace}
 \usepackage{ulem}
\usepackage{color}

\begin{document}

\preprint{APS/123-QED}

\title{Dynamic interfaces for contact-time control of colloidal interactions}

\author{Yaxin Xu}
\email{equal contribution}
\affiliation{Department of Chemical Engineering, University of California, Santa Barbara, Santa Barbara, CA, United States.}
\author{Kyu Hwan Choi$^*$}
\affiliation{Department of Chemical Engineering, University of California, Santa Barbara, Santa Barbara, CA, United States.}%
\author{Sachit G. Nagella}%
\affiliation{Department of Chemical Engineering, University of California, Santa Barbara, Santa Barbara, CA, United States.}%
\author{Sho C. Takatori}%
\email{stakatori@ucsb.edu}
\affiliation{Department of Chemical Engineering, University of California, Santa Barbara, Santa Barbara, CA, United States.}%

\date{\today}

\begin{abstract}
Understanding multibody interactions between colloidal particles out of equilibrium has a profound impact on dynamical processes such as colloidal self assembly. 
However, traditional colloidal interactions are effectively quasi-static on colloidal timescales and cannot be modulated out of equilibrium.
A mechanism to dynamically tune the interactions during colloidal contacts can provide new avenues for self assembly and material design.
In this work, we develop a framework based on polymer-coated colloids and demonstrate that in-plane surface mobility and mechanical relaxation of polymers at colloidal contact interfaces enable an effective, dynamic interaction.
Combining analytical theory, simulations, and optical tweezer experiments, we demonstrate precise control of dynamic pair interactions over a range of pico-Newton forces and seconds timescales.
Our model may be used to engineer colloids with exquisite control over the kinetics and thermodynamics of colloidal self-assembly dynamics via interface modulation and nonequilibrium processing. 
\end{abstract}

\maketitle


The material properties of colloidal suspensions depend on the multibody interactions between constituent particles\cite{Russel1989}.
These interactions may be programmed through functionalizing colloids with surface species such as DNA linkers \cite{Alivisatos1996, Macfarlane2019,Chen, Mirkin1996} or polymer brushes \cite{Biancaniello2005,Akcora2009} to guide or hinder colloidal aggregation.
In modelling such systems, one typically assumes a separation of timescales between the rapid relaxation of surface species and the colloidal Brownian diffusion \cite{Angioletti-Uberti2016} to obtain an effective, `static' pair potential, which solely depends on the instantaneous pair separation\cite{Loverso2012,Egorov2008}.
Although only exact at equilibrium, static potentials have been applied successfully to describe many colloidal suspensions out of equilibrium.

In some cases, however, nonequilibrium processes such as hydrodynamic flows \cite{Vermant2005} or kinetic arrest \cite{Segre2001} drive colloids together or apart faster than the surface species equilibration, resulting in a nontrivial interplay between the macroscopic process timescale and kinetics at the contact interface.
For instance, the stiffening of particle-particle contacts in dense colloidal suspensions can lead to logarithmic growth in the elastic moduli over time, in the absence of microstructural changes \cite{Bonacci2020, bonacci2022yield}.
Suspensions of polymer-grafted particles can exhibit shear thickening through hydrodynamic interactions and contact relaxation \cite{Melrose2004_1, Melrose2004_2}. 
Dynamical interactions are also biologically relevant; cell membranes are coated by receptors and biopolymers which spatially rearrange over cell-cell contact timescales of seconds to minutes to trigger T cell activation \cite{Kupfer1998,Davis2006}.
In these systems, a static potential is likely inadequate for predicting nonequilibrium pairwise interactions.
By modulating the intrinsic timescales at colloidal contacts, we aim to engineer a dynamic pair potential for multiscale control of colloidal interactions out of equilibrium.

\begin{figure}[hbt!]
\centering
\includegraphics[width=\columnwidth]{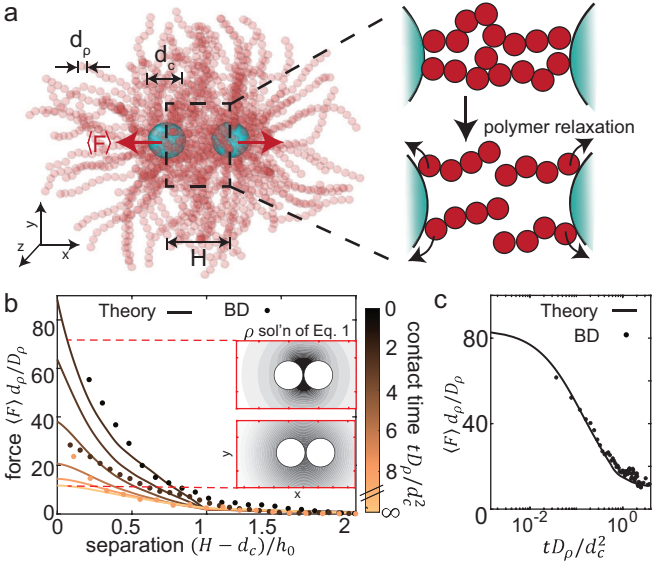}
\caption{\label{fig1}
Effective repulsive forces between polymer-grafted colloids decay as a function of colloid-colloid contact time due to polymer relaxation at the contact interface. 
(a) Brownian Dynamics (BD) simulation of two colloids (blue) with diameter $d_{\text{c}}$ at separation $H$ coated by same-length, end-grafted polymers (red) with surface-mobile grafting sites and bead diameter $d_{\rho}$. 
When the colloids are brought instantaneously into small $H$, polymers are forced into nonequilibrium configurations and generate a large effective force $\langle F\rangle$ between the colloids. 
As the polymers relax towards equilibrium, the effective interactions decay.
(b) Effective colloidal forces as a function of $H$ for short (black) to long (yellow) contact times.
Inset shows numerical solutions of Eq.~\ref{Eq1}-\ref{Eq2} for polymer density $\rho$ at short (top) and long (bottom) contact times. 
Dark regions indicate higher polymer density.
(c) Effective colloidal force as a function of contact time at colloidal contact, $H = d_{\text{c}}$.
Solid lines are solutions to Eq.~\ref{Eq1}-\ref{Eq2}, and markers are BD simulations. 
} 
\end{figure}

Consider the system in Fig~\ref{fig1}a: two colloids are coated by end-grafted polymers whose grafting sites are free to diffuse laterally along the surfaces. 
Colloids are brought to a small separation distance instantaneously and held fixed at those positions.
Shortly after contact, colloids experience a strong steric repulsion due to polymer overlap between opposing surfaces. 
However, through grafting-site diffusion and chain relaxations at longer times, the polymers assume configurations that lower their overall energy, thereby reducing the effective repulsion experienced by the colloids. 
This contact-time dependent interaction relaxes over colloidal timescales and can affect overall suspension dynamics.
Mechanistic understanding of these interactions has not been previously considered theoretically or experimentally. 

In this work, we combine theory, simulations, and experiments to directly measure the force transmission between two colloidal particles coated by surface-mobile polymers as a function of their contact time. 
We find that the relaxation timescale of this dynamic interaction is modulated by nonequilibrium protocols such as colloid approach speed.
Our understanding of dynamic pair interactions may be used to engineer colloids with precise control over nonequlibrium colloidal self-assembly.

\textit{Smoluchowski-based theory---}
We now provide an overview of our analytical theory to capture relaxation dynamics of surface-bound, semirigid polymers. 
The probability density $\rho(\textbf{h},t|H)$ of finding a monomer at position $\textbf{h}$, given two colloids of size $d_{\text{c}}$ at a separation $H$, satisfies the Smoluchowski equation:
\begin{equation}\label{Eq1}
\frac{\partial \rho}{\partial t} = -\nabla_{\textbf{h}} \cdot \textbf{j}
\end{equation}
where the flux contains thermal and interparticle contributions: 
\begin{equation}\label{Eq2}
\textbf{j} = - D_{\rho} \nabla_{\textbf{h}} \rho - D_{\rho} \rho \nabla_{\textbf{h}} V(\textbf{h}|H)/k_{\text{B}}T.
\end{equation}
Neglecting hydrodynamic interactions, $D_{\rho}$ is the Stokes-Einstein-Sutherland diffusivity of the monomer.
Assuming semidilute polymers, the interparticle potential $V = V_{\text{brush}} + V_{\text{HS}}$ is a sum of the entropic penalty of chain stretching and hard-core repulsions.
Eq.~\ref{Eq1} is numerically evaluated using the finite element software package FreeFEM++ \cite{Hecht2012}. 
Additionally, we perform coarse-grained Brownian Dynamics (BD) simulations using HOOMD-Blue (Supp.~Video 1-3) \cite{Anderson2020}.
Polymers are modeled with identical properties using a bead-spring model with semi-flexibility \cite{Kremer1990}, where
the grafting site is allowed to undergo diffusive translation along the surface (Fig.~\ref{fig1}a).
To quantify the effective colloidal interaction mediated by brushes of polymerization $M$ and surface density $n_{\rho}$, we compute the force $\langle F (\textbf{h},t|H) \rangle$ exerted by polymers on the colloids along their line of centers, where $F = - n_{\rho}M\partial_{H} V$ and the brackets $\langle ... \rangle = \frac{1}{2} \int \rho ... d\textbf{h}$. 

In Fig.~\ref{fig1}b, we plot the force exerted between the colloids as a function of the inter-colloidal separation $H$ for a family of contact times.
The inset shows cross-sectional monomer density solutions to Eq.~\ref{Eq1} - \ref{Eq2} for short and long contact times.
At a given separation, we observe that the repulsive forces decay as a function of contact time.  
At small times, $t \ll d_{\text{c}}^{2}/D_{\rho}$, we observe a repulsive force that strengthens when brushes are fully overlapped, $(H-d_{\text{c}})=h_{0}$, resulting from high osmotic pressure across the contact interface \cite{Russel1989}.
When the contact times exceed the diffusive timescale for the grafting site to explore the colloidal surface, $\tau_{\text{R}} \sim d_{\text{c}}^2/D_{\rho}$, polymers chains and their grafting sites have substantially depleted from the interfacial region, resulting in an order of magnitude decrease in force.
Unlike static pair potentials, this dynamic interaction is unusual because the colloids' instantaneous separations do not fully capture their force and stress transmission. 
We also note that this dynamic interaction is governed by the intrinsic timescales of the polymers and is distinct from externally-imposed, time-varying potentials \cite{Swan2012,Sherman2016,Tagliazucchi2016}. 

In the infinite time limit, compressed polymers fully relax through diffusive redistribution of their grafting sites out of the contact interface and spatial reorganization of the polymer chain.
In Fig.~\ref{fig1}c, we plot the colloidal force as a function of contact time when colloids are in contact at the closest separation, $H=d_c$.
We show that the force decays exponentially towards the equilibrium value, suggesting a characteristic relaxation timescale associated with polymer reorganization over the colloidal surface (Fig.~\ref{fig1}c).
This relaxed force is weaker than static repulsion between polymer brushes whose grafting sites are not laterally mobile \cite{Dolan1974,Russel1989, Milner1991,israelachvili2011intermolecular}.
We note good agreement between theoretical predictions and our BD simulations despite the simplicity of our polymer model.

\begin{figure}
\includegraphics[width=\columnwidth]{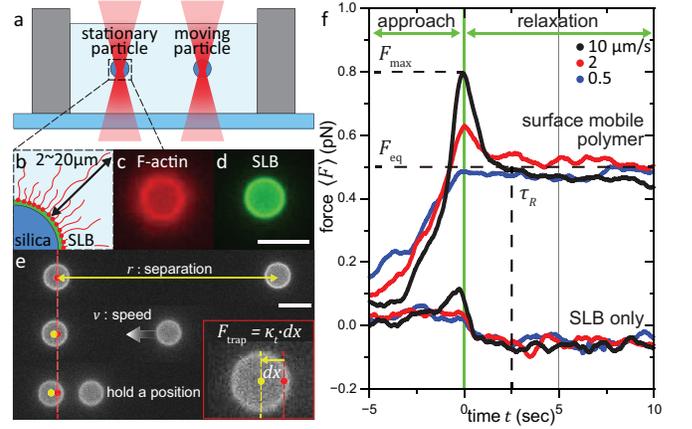}
\caption{\label{fig2}
Filamentous actin (F-actin) grafted on lipid bilayer-coated silica colloids generates contact-time dependent interactions between colloids.
(a) Experimental setup of optical laser tweezers and trapped particles in solution. 
(b) F-actin length ranges from 2$\mu$m to 20$\mu$m, with a mean $h_{0} \approx 5 \mu$m. 
(c,d) Fluorescence images of F-actin (red) bound to the lipid bilayer (green) containing polyhistidine tagged gelsolin and DGS-NTA(Ni).
(e) Force measurement method. 
Inset shows the displacement from laser focus (red dot) to the mass center of the stationary colloid (yellow dot). 
(f) Force as a function of time on beads with F-actin surface density $n_{\text{actin}}\approx 12,000/\mu \text{m}^2$ and a separate measurement for bilayer-only control. 
Solid lines are time-average curves with approach speeds of 0.5 $\mu$m/s (blue), $2\mu$m/s (red), and $10\mu$m/s (black). Times $t<0$ correspond to the approach step and $t \geq 0$ represent times when the colloids are at close contact. All scale bars are 5$\mu$m.
}
\smallskip
\end{figure}

\textit{Experimental measurement of pair interactions---}
So far, we have demonstrated that a nonequilibrium interaction exists between colloids coated with surface-mobile polymer layers through a theoretical model and BD simulations. 
Next, we present an experimental realization of this system and measure the interparticle interactions between polymer-coated colloids via optical tweezers (OT) \cite{Crocker1999}. 
Force measurements were conducted using the OT setup described in Fig.~\ref{fig2}a.
Two polymer-grafted beads were held in two separate traps focused more than 40$\mu$m from the bottom cover slip.
A supported lipid bilayer (SLB) containing dioleoyl-sn-glycero-3-phosphocholine (DOPC) was constructed on $d_{\text{c}}= 4\mu$m silica beads to enable mobility of surface species (Fig.~\ref{fig2}b,d) \cite{castellana2006solid}.
We chose filamentous actin (F-actin) as the grafted polymer for its ability to polymerize to large lengths \cite{sept1999annealing} and well-known mechanical properties \cite{claessens2006actin,brangbour2011force}. 
F-actin polymerization was quenched after reaching a length distribution of $2-20 \mu$m by washing out unreacted materials.
F-actin was end-grafted on the SLB by 6x-histidine tagged gelsolin to an anchoring lipid, 1,2-dioleoyl-sn-glycero-3-[(N-(5-amino-1- carboxypentyl)iminodiacetic acid)succinyl] (DGS-NTA(Ni)), which was doped in the bilayer over a range of $0 - 10\%$ to vary F-actin surface density between $n_{\text{actin}} \approx 0 - 12,000 / \mu \text{m}^2$ (Fig.~\ref{fig2}c, Supp.~Video 4, see SI for F-actin density characterization).
The mean separation between grafting sites is $10-20\text{nm}$, such that F-actin assumes a brush configuration.
1,2-dioleoyl-sn-glycero-3-phosphoethanolamine (DOPE) labeled with Atto-488 was added at 1\% for fluorescence.

In the experiment, a pair of colloids is placed in separate optical traps; one trap is stationary whereas the other trap translates at a fixed speed ($ v = 0.5-10 \mu$m/s) to bring the colloids from a large separation ($35 \mu$m) to a closest distance of $400-500$ nm before being fixed at this position for 20s (Fig.~\ref{fig2}e, Supp.~Video 5-6). 
We measured the stationary colloid displacement about its trap center at every time step, $dx$, following $\langle F \rangle = F_{\text{trap}}=\kappa_{\text{t}} \cdot dx$, with a trap stiffness $\kappa_{\text{t}} = 0.5 - 0.7 \text{pN}/\mu \text{m}$.
During the approach step, we did not observe convection-induced accumulation of F-actin to the rear of the colloid (Supp.~Video 6), indicating that hydrodynamic forces do not macroscopically perturb the polymer distribution.
 
In Fig.~\ref{fig2}f, we plot the interaction force against time for various approach velocities, where the translating trap stops motion at $t=0$. 
We observed that the repulsive forces increase as the two colloids approach for times $t < 0$ due to F-actin interactions with the opposing colloidal surface, and is maximized at the closest separation, $\langle F (t=0)\rangle = F_\text{max}$.
At the fastest approach velocity, $10 \mu \text{m/s}$ (black curve), the repulsive force relaxes from $F_\text{max}$ to the equilibrium force, $F_\text{eq}$, on an observable timescale, $\tau_R \approx 2.5 s$, consistent with literature values for F-actin spatial reorganization over the colloidal size, $\tau_R \sim 4 \mu \text{m}^2/D_a \approx 2.7 s $, where $D_a = 1.5 \mu \text{m}^2/$s is F-actin diffusivity in solution \cite{Janmey1986}. 
We therefore rationalize that the in-plane fluidity of the membrane surface enables an exquisite control over the reorganization of F-actin at the contact interface and the force transmission between the colloids.

At slow approach velocities, $0.5 \mu \text{m/s}$, the repulsive force between the colloids immediately equilibrates --- their interactions are quasi-static because the polymers have sufficient time to reorganize during every step of approach.
This equilibrium force is related to a potential of mean force, $F_{0.5\mu \text{m/s}} = - \int \rho_{\text{eq}} \partial_{\text{H}} V d\textbf{h}$ (Fig.~\ref{fig1}b) where $\rho_{\text{eq}} = e^{-V/k_{\text{B}}T}$ is the equilibrium monomer distribution.
As a control, we show that the forces between SLB-coated colloids without F-actin remained approximately zero throughout, except for the small peak associated with a lubrication force at the largest velocity. 
The small negative $F_\text{eq} \approx 50\text{fN}$ indicates a weak, van-der Waals-type attractions.

Figure ~\ref{fig2}f is an experimental realization of our simulations in Fig.~\ref{fig1}b, where two colloids placed quickly into contact experienced a repulsive force that decays with contact time.
Using our membrane-coated colloids with different surface conditions, one can create a range of designer pair potentials with tunable contact-time interactions, as demonstrated theoretically in Fig.~\ref{fig1}.

\begin{figure}
\includegraphics[width=\columnwidth]{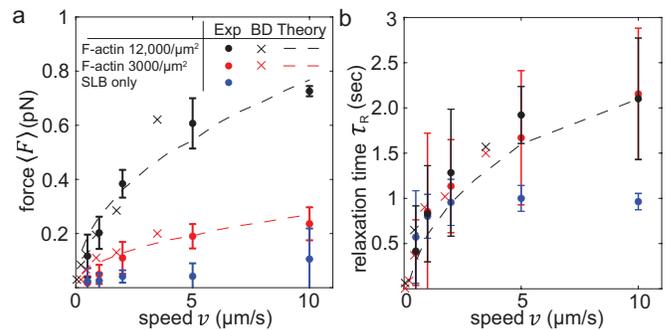}
\caption{\label{fig3}
Nonequilibrium process timescale modulates strength and relaxation of dynamic colloidal interactions. 
(a) Effective colloidal force versus approach speed at a center-center separation of 8$\mu$m, for large actin density ($12,000 / \mu \text{m}^2$) (black), moderate actin density ($3,000 / \mu \text{m}^2$) (red), and SLB only (blue).
(b) Relaxation time from peak force versus approach speed.
Dashed curves are solutions to Eq.~\ref{Eq1}-\ref{Eq2}, filled circles are experiments, and crosses are BD simulations. 
}
\smallskip
\end{figure}

The contact-time dependent interactions in Fig.~\ref{fig1}-\ref{fig2} arise from the nonequilibrium distributions of interfacial polymers. 
Therefore, any process that moves the colloids in and out of contact on a timescale that competes with polymer relaxation, such as hydrodynamic fluid flows and other non-conservative body forces, can induce a dynamic interaction.
To understand the impact of these competing timescales, we systematically varied the approach velocities of the colloids leading to their closest separation.

In Fig.~\ref{fig3}a, we measured the effective force as a function of approach velocity at a fixed colloidal separation ($H = 8 \mu$m) for two actin surface densities,  $12,000/\mu\text{m}^2$ and $3,000/\mu\text{m}^2$. 
We observe that the effective colloidal force increases for higher surface densities, consistent with our hypothesis that the polymer-mediated repulsion is induced by increased osmotic pressure (Fig.~\ref{fig1}b, inset).
Also, the forces generally increase for higher approach speeds, which we attribute to the degree of F-actin compression at the contact interface.
During a ``fast'' approach ($v > 2 \mu$m/s), F-actin of mean height $h_{0}\sim 5\mu$m is compressed at a timescale $\tau_{\text{process}} \sim h_{0} / v = 2.5$s, which is comparable to the F-actin reorganization on the colloid surface.
Thus, polymers compress without having sufficient time to explore favorable configurations.
Higher approach speeds induce an increasingly dense layer of interfacial F-actin, generating stronger forces. 
This repulsion begins to plateau at the highest approach speed ($10 \mu$m/s), possibly because polymers cannot infinitely accumulate.
Note in Fig.~\ref{fig3}a that the theoretical polymer configurations were generated for the initial approach (see SI for polymer initialization).

Our results confirm that faster approach processes drive polymers further away from their equilibrium distribution. 
Therefore, the approach timescale should not only influence the strength of polymer-mediated interactions but also control their relaxations toward equilibrium.
In Fig.~\ref{fig3}b, we plot the characteristic relaxation time $\tau_{\text{R}}$ of the effective force as a function of approach speed.
We observe that the relaxation time increases with faster approach speeds, suggesting that polymers equilibrate more slowly when strongly compressed.

Interestingly, the relaxation time is independent of the F-actin surface density (Fig.~\ref{fig3}b).
From our flux expression (Eq.~\ref{Eq2}), we conclude that it is the \textit{gradients} in polymer concentration along the colloidal surfaces, $\nabla_{\textbf{h}} \rho$, which drive relaxation towards equilibrium.
This is reminiscent of Marangoni forces that drive surfactant molecules from high to low concentrations \cite{Leal2007}.
Such a trend supports our theoretical framework of modeling relaxation as a diffusion-mediated process, as opposed to other mechanisms that depend on polymer concentration. 

In Fig.~\ref{fig3}, we have found that a free-draining model sufficiently captures the key physics behind our experimental trends. 
In general, however, hydrodynamic effects cannot be neglected. 
As a control, we show that forces between SLB-coated colloids scale linearly with approach speed, consistent with low-Reynolds number hydrodynamics (Fig.~\ref{fig3}a, Supp.~Video 7).
Moreover, lubrication theory predicts a constant relaxation time $\tau_{\text{SLB}} = 3\pi\eta d_{\text{c}}^2/(2\kappa_{\text{t}} (H-d_{\text{c}})) \approx 1$s, in agreement with experimental results (Fig.~\ref{fig3}b).
A more detailed discussion of the fluid-mediated forces may be found in the SI.

\textit{Discussion.---}
In this paper, we have demonstrated that functionalizing colloidal surfaces with laterally-mobile, end-grafted polymers generates a dynamic pair force which relaxes as a function of colloidal contact times. 
We observe that timescales of nonequilibrium processes driving colloids into contact non-trivially compete with the timescale of polymer brush reorganization at the contact interface.
Although more effects, such as rotational motion, frictional forces \cite{Mari2014,Lin2015}, and fluid-mediated forces \cite{Klein1993,Fredrickson1991,Doyle1996}, could influence the effective interaction, we have obtained proficient agreement between our Smoluchowski theory with BD simulations and OT experiments.
We believe our simple framework captures the essential nonequilibrium physics of polymer-mediated forces and relaxation at colloidal contacts.

\begin{figure}
\includegraphics[width=\columnwidth]{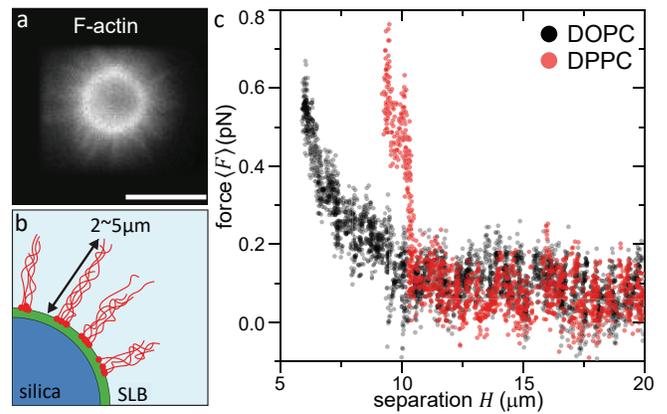}
\caption{\label{fig4}
Less-mobile surface-grafted F-actin spontaneously organizes into rigid bundles, inducing steep and repulsive interactions which buckle when colloids are brought together.
(a) Fluorescence image of end-grafted F-actin bundles coating the colloid whose SLB contains DPPC lipids and 10\% DGS-NTA(Ni).
(b) Schematic of bundled F-actin with a mean bundle length $2\mu$m, whose grafting sites are immobile, non-rotating, and phase-separated on the SLB surface. 
(c) Effective force changing in colloidal separation $H$ when actin-coated colloids approach at 0.5$\mu$m/s. 
F-actin anchored to less-mobile DPPC SLB (red) mediates sharp force increases and buckling near $H=10\mu$m, in contrast to monotonic repulsion when anchored to the more-mobile DOPC SLB (black).
}
\smallskip
\end{figure}

We conclude this Letter by observing that surface chemistry and composition may be leveraged to engineer different types of contact-time dependent interactions.
As a demonstration, we synthesize F-actin-coated colloids whose bilayers contain 1,2-dipalmitoyl-sn-glycero-3-phosphocholine (DPPC), which forms more rigid membranes compared to DOPC lipids \cite{Rawicz2000}. 
Interestingly, the DPPC membrane organizes F-actin into rigid, protruding bundles with immobile and non-rotating grafting sites (Fig.~\ref{fig4}a,b, Supp.~Video 8).
We surmise that micro-phase separation on the bead surface \cite{chen2014three} induces gelsolin to form small patches, thereby organizing F-actin into bundles.

In Fig.~\ref{fig4}c, we perform OT experiments to compare the effective colloidal force between DOPC and DPPC membrane conditions as a function of colloidal separation at a fixed approach speed ($0.5\mu$m/s).
Unlike DOPC colloids, we observed sharp force increases and buckling when F-actin bundles on DPPC colloids begin to overlap, $H = 10\mu$m (Fig.~\ref{fig4}c).
Because F-actin organizes into stiff bundles, we note that these DPPC surfaces are very different from the DOPC surfaces, and their interaction dynamics cannot be directly compared.
Mechanistic understanding of the force transmission between bundle-forming F-actin layers is left for future work.

More generally, our conceptual framework of contact-time dependent interactions is applicable to systems beyond pair interactions of lipid-coated particles.
For example, the interactions of a third colloid to a dimer would depend on surface rearrangement of mobile species. 
By extending to N-particle interactions, we can engineer the kinetics and morphology of multi-body assemblies.
Our framework is also applicable to multi-component interfaces with adhesive linkers and repulsive brushes, analogous to ligand-receptor binding at crowded cell-cell junctions, and allows us to dynamic tune between repulsive and attractive interactions. 
Furthermore, the timescale competition between hydrodynamic shear and dynamic pair interactions may impact particle suspension rheology \cite{Wagner1990}.
Finally, our framework may help understand other complex dynamic interfaces such as surfactant-laden emulsions \cite{Cristini1998,Chesters2000}, colloids coated by polymers with adsorption and desorption rates \cite{Pincus1984,Pefferkorn1990}, cell surfaces where proteins undergo lateral rearrangement upon cell-cell contact \cite{Kupfer1998}, and uptake of macromolecules on membranes with characteristic wrapping times \cite{Nel2009}. 

\begin{acknowledgments}
We would like to thank Fyl Pincus, Eric Furst, Tim Lodge, Glenn Fredrickson, and Todd Squires for valuable discussions and feedback. 
This material is based upon work supported by the Air Force Office of Scientific Research under award number FA9550-21-1-0287.
Y.X. acknowledges support from the Dow Discovery Fellowship at UC Santa Barbara. 
S.G.N. acknowledges support from the National Institutes of Health (1T32GM141846). 
S.C.T. is supported by the Packard Fellowship in Science and Engineering.
Acknowledgment is made to the Donors of the American Chemical Society Petroleum Research Fund for partial support of this research.
Use was made of computational facilities purchased with funds from the National Science Foundation (OAC-1925717) and administered by the Center for Scientific Computing (MRSEC; NSF DMR 1720256) at UC Santa Barbara.
\end{acknowledgments}

%

\end{document}